# whyqd: Auditable and reusable crosswalks for fast, scaled integration of scattered tabular data

Gavin Chait

Whythawk, France, and Centre for Advanced Spatial Analysis, UCL, UK

gchait@whythawk.com or g.chait@ucl.ac.uk

**ABSTRACT**

*This paper presents whyqd (/wɪkɪd/), an open-source curatorial toolkit intended to produce well-structured and interoperable data. Curation is divided into discrete components, with a schema-centric focus for auditable restructuring of complex and scattered tabular data to conform to a destination schema. Task separation allows development of software and analysis without source data being present. Transformations are captured as high-level sequential scripts describing schema-to-schema mappings, reducing complexity and resource requirements. Ultimately, data are transformed, but the objective is that any data meeting a schema definition can be restructured using a crosswalk. whyqd is available both as a Python package, and as a 'no-code' visual web application. A visual example is presented, derived from a longitudinal study we manage where scattered source data from hundreds of local councils are integrated into a single database.*

## 1   INTRODUCTION

Spreadsheets are the most common form of data storage and exchange for the public and private sector. Lack of standardisation in nomenclature, especially in highly complex organisations, often produces hard to reconcile data structures. Before these diverse, poorly-structured and scattered tabular data can be reused, they must first be transformed to conform to a standardised structural schema. Time and complexity for transformation at scale is a major obstacle to discovering whether these data are useful in the first place.

Current data tidying toolkits are data-centric, promoting workflows where curators restructure data directly at row and column level, potentially interacting directly with database environments. This is labour- and skill-intensive, and often accomplished through time-consuming development of data structuring scripts and source code, and often sensitive to small format changes.

This migrates complexity from tabular source data to the restructuring script.

The more time required to read, revise and review code and scripts developed to transform semi-structured data, the less likely it is that these scripts will be reused to transform other data, and the more they present a new obstacle to data interoperability.

**whyqd** (/wɪkɪd/) is an open-source schema-focused curatorial toolkit intended to produce well-structured and interoperable data for research analysis. Its objective is not to explore data, or perform interim analysis, but support schema-based workflows and validations, and documented and auditable transformations.

whyqd was developed in response to our curatorial needs for our openLocal.uk project [Chait, 2016], a quarterly-updated commercial location database, aggregating open data on vacancies, rents, rates and ratepayers, into an integrated time-series of individual retail, industrial, office and leisure business units.

openLocal represents a worst-case example of integrating scattered data.

Every three months, we import semi-structured spreadsheets from 250 to 300 local councils across the UK. These need to be restructured to conform to a single schema, categorical data terms redefined to an ontology, and then data cleaned and validated.

In a context often found in large organisations, complexity arises from failure of local councils to develop common data management standards, and an absence of any agreed schema. Our source data can be in any tabular format (XLS, XLSX and CSV).

This lack of consistency is often exacerbated by high staff turnover in public- and private organisations, which leads to loss of domain specific knowledge and may result in regular, undocumented data structure changes in subsequent releases.

This paper encapsulates our approach to data management and organising data interoperability at scale. It starts with a reflection of current data transformation practice and the challenges of data-centric curation, presents whyqd's architecture and process for intentional data transformation, and demonstrates how whyqd works with a visual case study taken from our work on openLocal.

## 2 APPROACHES FOR TIDY DATA

### 2.1 Curation and Tidy Data

Ideally, source data are **tidy** [Wikham, 2014]; well-structured, conforming to standard data types, and straightforward to convert to alternative forms. More typically, source data were designed for presentation or one-time use, not reuse, and the task of curation for interoperability is significantly more challenging (Figure 1).

| HDI rank | | Total population (millions) | | | Annual population growth rate (%) | | Urban population[a] (% of total) | | |
|---|---|---|---|---|---|---|---|---|---|
| | | 1975 | 2005 | 2015[b] | 1975-2005 | 2005-15[b] | 1975 | 2005 | 2015[b] |
| **HIGH HUMAN DEVELOPMENT** | | | | | | | | | |
| 1 | Iceland | 0.2 | 0.3 | 0.3 | 1.0 | 0.8 | 86.7 | 92.8 | 93.6 |
| 2 | Norway | 4.0 | 4.6 | 4.9 | 0.5 | 0.6 | 68.2 | 77.4 | 78.6 |
| 3 | Australia | 13.6 | 20.3 | 22.4 | 1.3 | 1.0 | 85.9 | 88.2 | 89.9 |
| 4 | Canada | 23.1 | 32.3 | 35.2 | 1.1 | 0.9 | 75.6 | 80.1 | 81.4 |
| 5 | Ireland | 3.2 | 4.1 | 4.8 | 0.9 | 1.5 | 53.6 | 60.5 | 63.8 |

*Figure 1: UNDP Human Development Index 2007-2008, human-readable spreadsheet [green triangles indicate numbers stored as text as a result of included non-numeric data]*

Data **curation** includes all the processes and techniques needed for ethical and reproducible data creation, management, transformation, and presentation for reuse.

Curation must ensure **data probity**:

- Identifiable and uniquely verifiable input source data,
- Transparent methods for restructuring of source data into data used in analysis,
- Accessible restructured data used to support research conclusions,
- A repeatable, auditable transformation process which produces the same data.

Researchers may disagree on research conclusions or analysis. They should not have cause for disagreement on the probity of underlying data used to produce that analysis.

A **crosswalk** is any set of steps used to transform tabular source data into a destination structure. Individual steps are mappings of relationships between source and destination fields defined in each metadata schema. Ideally, these are one-to-one, where a field in one has an exact match in the other. In practice, it's more complicated than that.

A **schema** defines the structural organisation of tabular data. Each column is identified by a **field** name and defined by conformance to technical specifications (such as datatype, e.g. text, integer, date, etc.). These, along with field constraints and sensible defaults, support interoperability.

**2.2 Current Methods for Data Transformation**

Curators currently directly explore tabular source data, and implement a transform in as close to real-time as possible.

**Microsoft Excel** is a commonly-used spreadsheet application for manual data restructuring. Manual transforms are inherently **destructive**, as the underlying processing steps are typically not recorded or performed in the same way by different curators. As example, reviewing changes or auditing if errors or data loss were introduced.

Pandas, Tidyr, OpenRefine and Trifacta are software for **non-destructive** curatorial data manipulation, with methods for managing and saving reusable scripts for data transforms.

**Pandas** [McKinney, 2010] and **Tidyr** [Wickham, 2014] arose out of a dedicated and intentional effort to work with diverse data of any scale. They are programmatic, and require knowledge of the data being transformed and of their respective coding languages (Python and R). Both allow visual display of tabular data, but these are snapshots of the data state. You do not interact with the tabular display, unlike spreadsheets.

Tidyr is more sophisticated, in that common transforms are more readily available as standardised language, while Pandas requires greater effort, both require a degree of programming knowledge and experience. Given that not every data curator can code, or code in these specific languages, two visual approaches to data tidying were developed.

**OpenRefine** is an open-source toolkit running on a curator's computer (Figure 2).

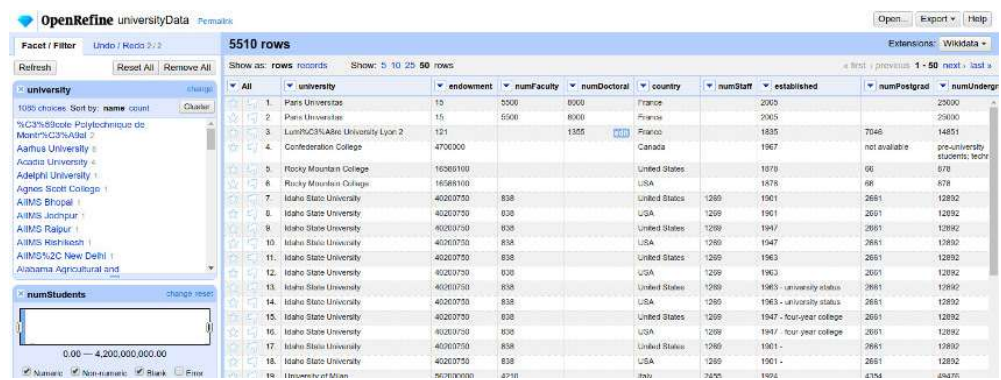

*Figure 2: OpenRefine visual interface*

**Trifacta** (originally Potter's Wheel, now Alteryx) is a cloud-based approach to support collaboration and offloading transformation tasks to dedicated servers. It is more supportive of less technical curators, and tries to identify and correct common errors automatically [Raman, 2001] (Figure 3). It runs in a browser, and curators interact with an abstracted subset of the entire dataset, and not the dataset itself [Raman, 1999].

Both OpenRefine and Trifacta are, effectively, call-and-response systems in which the curator defines changes, sees the impact of these changes on their data on screen in real-time, and can review the script of transformations as they are created. Both mimic Excel in displaying tabular data and encouraging a data-centric approach to curation.

Scripts are intended for reuse. Trifacta, being server-based, can operate autonomously.

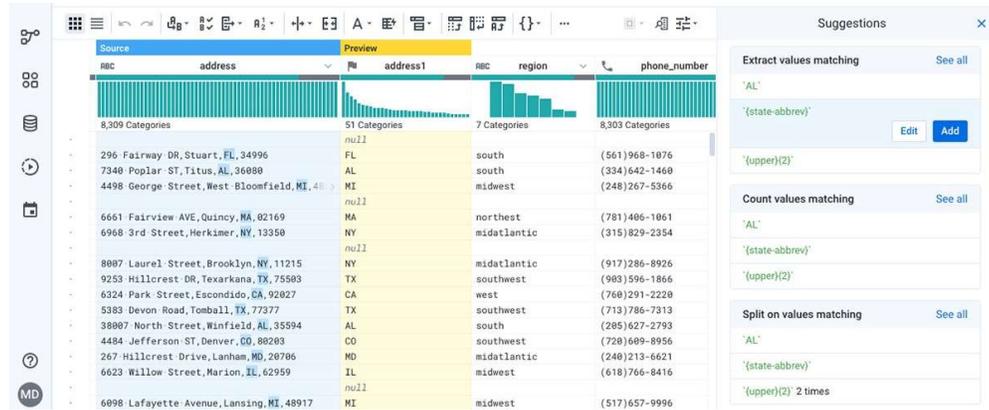

*Figure 3: Trifacta visual interface*

## 2.3 Risks with Granular Data-Centric Curation

These approaches assume that crosswalks must be performed on data, yet there are two discrete objectives during transformation:

1) **Restructuring** source columns and types into destination columns and types,
2) **Validation** and correction of values to conform to the types for each column.

Validation and correction of data are usually performed on individual values or rows, and often requires custom operations. Tidyr and Pandas are already coded, and OpenRefine and Trifacta permit the curator to "escape" their standard transform grammar and write code to perform complex row and value-level changes. These are added to the crosswalk script, but this specificity comes at the expense of simplicity, clarity and reusability.

Reviewing these complex, non-standardised scripts for accuracy is difficult. Reusing these scripts – especially where row-level changes were made – only works if data share identical column- **and** row-level structure. Complex data are traded for complex scripts.

Restructuring this bespoke process into a defined and intentional series of components would permit each to be organised, optimised, and appropriate toolkits to be designed.

## 3   WHYQD PROCESS FOR INTENTIONAL DATA TRANSFORMATION

whyqd (/wɪkɪd/) is an open-source curatorial toolkit intended to produce well-structured and interoperable data for research analysis. It supports trust in research by ensuring complete and unambiguous probity in the curation of all source data.

whyqd's focus is on auditable restructuring of complex and scattered tabular data to conform to a single destination schema. Validation is supported, but not its purpose. It is available both as a Python package, and as a "no-code" visual web-based application.

### 3.1 Source Data and the Minimum Transformable Schema

*Crosswalks are developed on schemas, not source data.*

whyqd captures crosswalks as schema-to-schema mappings, rather than data-to-data mappings. This reduces complexity for curators and software since fewer resources are required. Ultimately, data are transformed, but the objective is that any data meeting a particular schema definition can be restructured using a crosswalk.

Data validation and correction are ordinarily left to downstream systems. All whyqd requires of a source schema is that it have uniquely addressable fields, and that it preserves all data without requiring knowledge of specific data types.

This requirement is the **minimum transformable schema** which is derived from source data with the least assumptions possible, and according to these rules:

- For every column in the data source, there must be a matching field in the schema. Each column is either defined in the data's header row, or as a separate, ordered list of header names in the schema which match the listed fields.
- Strings are default field types to ensure source data preservation. There is no type guessing, and no potential for ambiguity during transformation. Any non-string type must be specifically assigned by the curator and documented in the schema.
- Categorical terms must be defined in the schema, and not during transformation.

If there is no header row, or the header row is complex (such as a combination of merged cells across multiple rows), the curator can instruct whyqd to define unique referenceable fields or accept a new set of field names on load.

Crosswalks are defined in terms of a destination schema, allowing software and analysis to be developed independently on synthetic data, without source data being present. Data validation is a separate process, permitting encapsulation of each component of a project.

### 3.2 Action Language Scripts as Crosswalk Documentation

*Crosswalks are read more than they are written.*

whyqd reduces crosswalks to a series of **action** scripts, each defining an individual step which must be performed to restructure source- into a destination schema. Scripts are written as a text string conforming to a standardised template, summarised as "Perform this action to create this destination field from these source fields.":

```
"ACTION > 'destination_field'::'destination_term' < 'source_term'::['source_field', 'source_field']"
```

Reading a script from left-to-right, from destination to source, goes from destination schema terms a curator can be expected to know, to less familiar source schema terms.

A set of standard actions permit complex crosswalk mappings:

| ACTION | > FIELD | > TERM | < TERM | < FIELD | < ROWS |
|---|---|---|---|---|---|
| CALCULATE | X | | | [m X,] | |
| CATEGORISE | X | X | [X,] | X | |
| COLLATE | X | | | [X, m,] | |
| DEBLANK | | | | | |
| DEDUPE | | | | | |
| DELETE_ROWS | | | | | [X,] |
| NEW | X | | [X] | | |
| PIVOT_CATEGORIES | X | | | X | [X,] |
| PIVOT_LONGER | [X, X] | | | [X,] | |
| RENAME | X | | | [X] | |
| SELECT | X | | | [X,] | |
| SELECT_NEWEST | X | | | [X m X,] | X |
| SELECT_OLDEST | X | | | [X m X,] | |
| SEPARATE | [X,] | | X | [X] | |
| UNITE | X | | X | [X,] | |

*Table 1: Alphabetical list of whyqd actions, with required terms and format.*

- **Fields** are the schema field names, while **terms** are context-dependent on the action, but usually a specific source data value or character from the field values.
- **Square brackets []** are used exactly as shown, where required.
- **X** stands for a field or term, and **m** for an action-specific term modifier,
- X requires only a single term, [X] a single term inside square brackets,
- [X, X] only two terms accepted,
- [X,] any number of terms,
- [m X,] any number of terms, but each term requires a modifier,
- [X m X,] any number term-term relationships related by a modifier,
- [X, m,] any number of terms or modifiers, in any combination.

As an example, `CATEGORISE` the source data values (terms) `Sports Club (Registered CASC)` and `Mandatory` from the `Current Relief Type` column (field) as the term `other` in the destination field `occupation_state_reliefs`:

```
"CATEGORISE > 'occupation_state_reliefs'::'other' < 'Current Relief Type'::['Sports Club (Registered CASC)', 'Mandatory']"
```

Each of these actions are explained, with examples, in the documentation [Chait, 2024].

### 3.3 Schema-to-Schema and Data-to-Schema-to-Schema Crosswalks

*Explicit is better than implicit. Don't leave transformation steps out.*

The objective of whyqd is not to explore data, or perform interim analysis, but to support schema-based workflows and validations, and documented and auditable transformations.

A whyqd curation process can be split into discrete, specialised tasks:

- **Data acquisition** from scattered sources,
- **Source file import** and hashing for probity using BLAKE2b [Aumasson, 2013],
- Derivation of **minimum transformable schemas**,
- **Labelling of schemas** to allocate field types, or derive categorical data,
- Creation of appropriate **crosswalk scripts**.

Where source data conforms to a known schema-to-schema crosswalk then it may simply be reused. With time, as a library of such crosswalks are created for a specific data integration project, less time is spent developing new crosswalks.

## 4 WHYQD SOFTWARE IMPLEMENTATION AND CASE STUDY

> This section is both a description of the underlying software, and a case study taken from direct experience. Coded examples are available in the Python package documentation [Chait, 2024], which extensively covers combining diverse data sources, dealing with complex, merged header rows (such as that presented in Figure 1), badly formatted data types, and misaligned columns.
>
> This case study will focus on the visual 'no-code' web application and demonstrate an approach to rapid, transparent and collaborative development of crosswalks. It tries to strike a balance between technical overview and definitions, and the real-world process of producing data transformations at scale.

Our openLocal.uk project [Chait, 2016] is used by researchers and the UK government but, during COVID in 2020-21, these data were part of the research base informing impact assessment for the various lockdowns and the economic recovery afterwards.

Our objective is a sequential event-based data series, capturing everything that happened for every commercial address in a particular source from a local council.

### 4.1 Python and Modin to Transform Millions of Rows

whyqd is a wrapper around Modin, a Python package based on Pandas. Switching from Pandas to Modin allows the use of a process management package called Ray, which gives the curator the ability to specify the number of threads they wish to use, and the memory they wish to allocate to each thread.

whyqd crosswalks are composed of action scripts that perform what would ordinarily be repetitive, but complex, code. It consists of discrete functionality to support a workflow, and has a **parser** – for interpreting action scripts, and validating them against the schemas – and a **transformer** for implementing the script to produce transformed data.

### 4.2 Define a Destination Schema

Schemas are defined as a JSON Schema-compliant format, with source data complying with Table Schema [Walsh, 2012]. They can be exported and saved as JSON files.

*Figure 4: whyqd, schema field creation showing a subset of its definitions and constraints*

Individual schema fields require only a name and type, but additional constraints can also be defined, especially when using categorical ontologies.

**4.3 Source Data and Deriving a Minimum Transformable Source Schema**

In whyqd, a **project** is assigned a **schedule** of data processing **tasks**, with consequent **activities** resulting in data transformation **resources**, and with a destination **schema**.

*Figure 5: a whyqd project, with links to its schedule, activities and schema, and a corresponding task, with links to import, its source, and current activities and state*

The curator can work in a flow suitable to them, usually batching import tasks before moving on to the next steps.

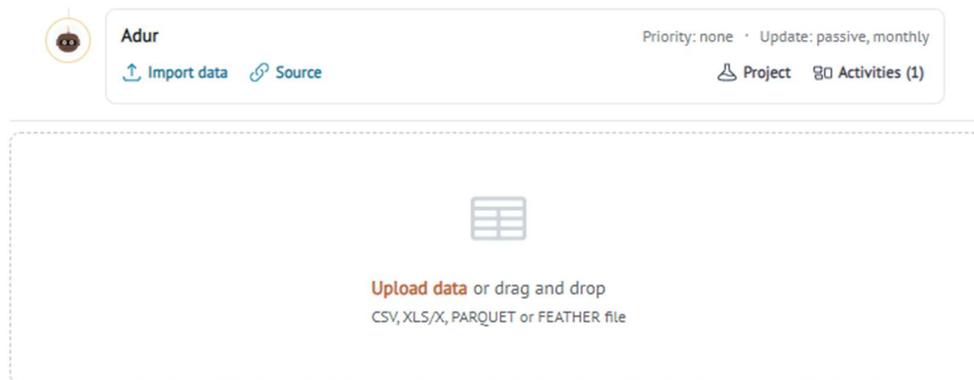

*Figure 6: whyqd import dialogue for a scheduled task*

On import, whyqd automatically processes the source data and derives its minimum transformable schema according to the rules stated above (Figure 7). Sources can be Excel (XLS and XLSX), CSV, Parquet or Feather. If of multiple sheets, whyqd creates independent resources for each sheet. More complex CSV encodings are managed by including Pandas import parameters in the import script [Pandas, 2020].

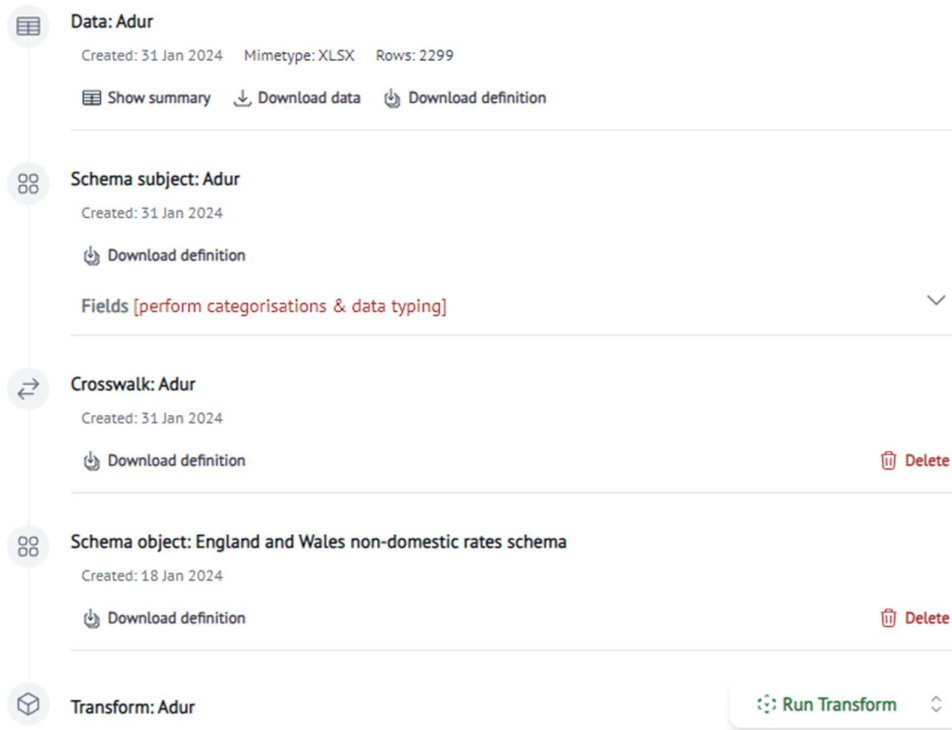

*Figure 7: a whyqd resource, with source data, subject and object schemas, and crosswalk*

If the combination of source and destination schemas match an existing crosswalk, it will automatically be assigned, and authority handed to the curator to validate further.

There is scope for performing introspection on existing crosswalks to compile a "glossary" of field-to-field crosswalks independently of schema definitions, but this has not been implemented and presents a risk of overfitting inappropriate matches.

The curator can review the derived schema and modify as required, from applying data types, to further deriving or applying categorical types onto source fields.

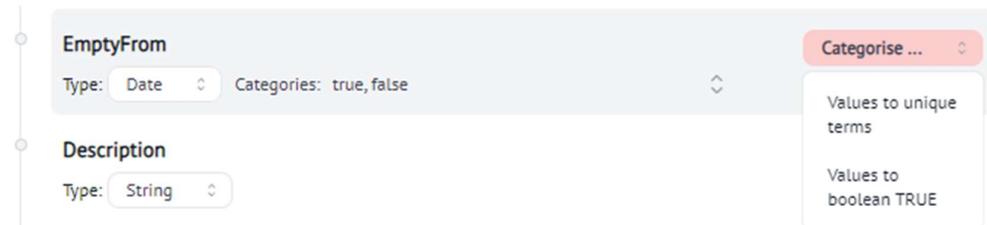

*Figure 8: deriving categories for values in source columns can be as unique terms, or the presence or absence of any value (Boolean) – the type, as a date, has also been set*

In this case study, there is no empty state field corresponding to the empty date. We can use the dates in `EmptyFrom` as an indicator by assigning boolean `True` / `False` values to the presence or absence date entries (Figure 8), then creating a new column with these values.

As example:

```
"CATEGORISE > 'occupation_state'::True < ' EmptyFrom'::[True]"
"RENAME > 'occupation_state_date' < ['EmptyFrom']"
```

Here, we used `EmptyFrom` twice, to create a categorical field, and to remap the date field.

The curator can also, as with Pandas or Tidyr, review a static data extract (Figure 9).

*Figure 9: a static sample of the first 50 rows of the source data*

### 4.4 Defining crosswalks

whyqd presents a visual drag 'n drop / 'no-code' interface for the curator (Figure 10), with an all-in-one workspace including schema highlighting, data extracts and crosswalk.

No-code, here, refers to the assistance provided by the app. Dragging an action into the workspace from the list of actions creates the appropriate action structure, with responsive drop-down lists for selecting fields. Validated actions highlight the related lists of source and destination schema fields as having been remapped.

The sequence of actions can be reordered and validated, and the workspace can be shared and reviewed by others to ensure consensus.

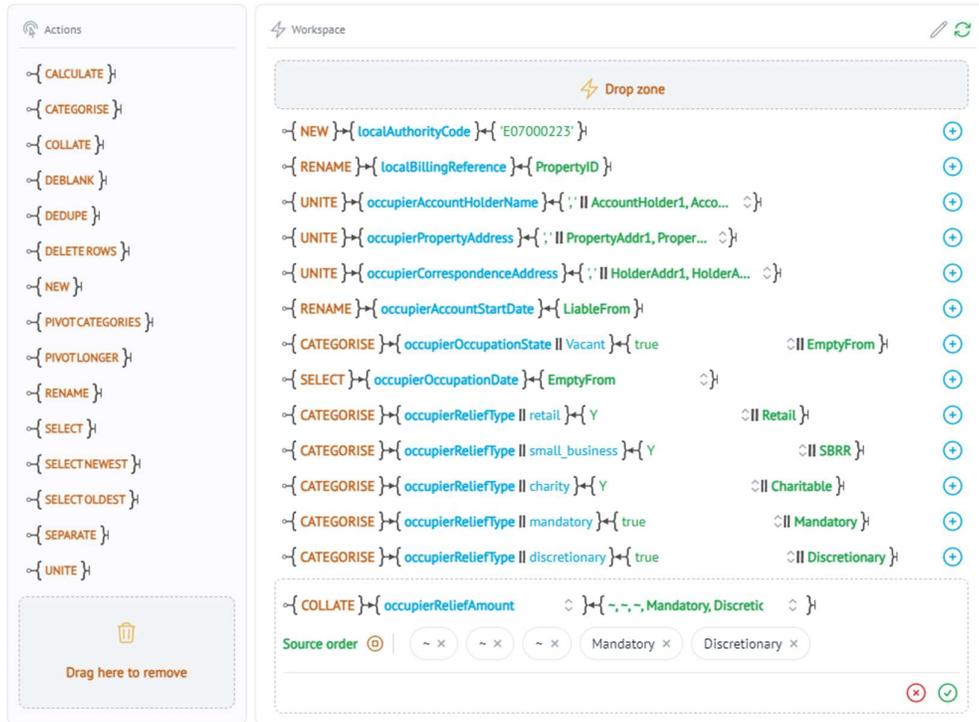

*Figure 10: whyqd crosswalk workspace showing the script and editing interface*

The crosswalk can be exported as a JSON file and read as easily as in the web app, or shared for programmatic reuse:

```
NEW > 'localAuthorityCode' < ['E07000223']
RENAME > 'localBillingReference' < ['PropertyID']
UNITE > 'occupierAccountHolderName' < ',','::['AccountHolder1','AccountHolder2']
UNITE > 'occupierPropertyAddress' < ',','::[
    'PropertyAddr1', 'PropertyAddr2', 'PropertyAddr3',
    'PropertyAddr4', 'PropertyPostcode'
]
UNITE > 'occupierCorrespondenceAddress' < ',','::[
    'HolderAddr1','HolderAddr2','HolderAddr3','HolderAddr4','HolderPostcode'
]
RENAME > 'occupierAccountStartDate' < ['LiableFrom']
CATEGORISE > 'occupierOccupationState'::'Vacant' < 'EmptyFrom'::[True]
SELECT > 'occupierOccupationDate' < ['EmptyFrom']
CATEGORISE > 'occupierReliefType'::'retail' < 'Retail'::['Y']
CATEGORISE > 'occupierReliefType'::'small_business' < 'SBRR'::['Y']
CATEGORISE > 'occupierReliefType'::'charity' < 'Charitable'::['Y']
CATEGORISE > 'occupierReliefType'::'mandatory' < 'Mandatory'::[True]
CATEGORISE > 'occupierReliefType'::'discretionary' < 'Discretionary'::[True]
COLLATE > 'occupierReliefAmount' < [~,~,~,'Mandatory','Discretionary']
```

### 4.5 Transformations with Crosswalks

Once completed to the curator's satisfaction, the crosswalk can be run to produce transformed data in any of the available formats (Figure 11).

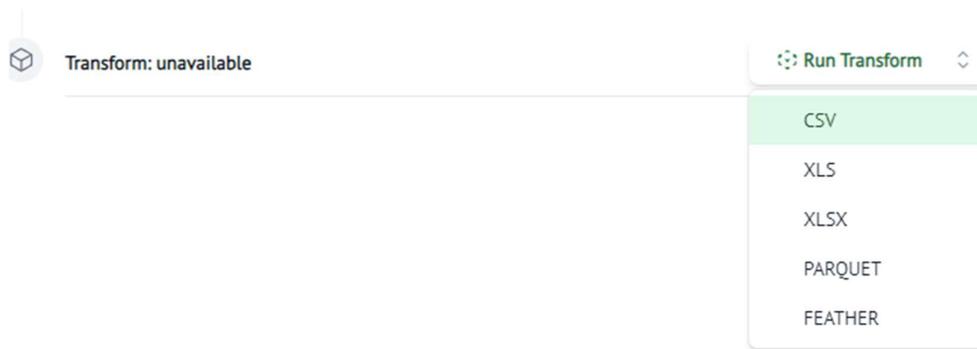

*Figure 11: Transform data into any of Excel (XLS and XLSX), CSV, Parquet or Feather*

All schemas, crosswalks, and transforms can be exported and shared.

These data are also available in bulk via the web application's API so that completed tasks can be imported automatically and processed for analysis elsewhere.

*Figure 12: whyqd display of extract of transformed data*

## 5   CONCLUSIONS AND FUTURE WORK

A critical risk in any data transformation arises from opaque destructive changes. Whether from manual spreadsheet restructuring or via complex coded scripts, lack of a formal language to describe and implement transformations may introduce errors into the transformed data or make audits unviable.

Transforming scattered data for interoperability is an expert task, requiring a combination of complex skills and *ad hoc* judgement calls. Assessing whether these decisions work may only be known once source data are integrated and used. At that stage, the ease with which changes or corrections may be made and implemented may present a significant challenge to improving the efficiency and success of a research process.

whyqd advantages this 'conversational' approach to data transformation. While it may provide little efficiency gains for low-volume data projects, as the diversity and complexity of source data grows, and greater reliance is imposed on the knowledge bank of existing crosswalks, it improves transparency, collaboration and reuse.

Our approach reduces transformation risk through formal scripted crosswalks which implement a repeatable, auditable transformation process and which can be used at scale.

whyqd promotes intentional and collaborative data transformation, which permits task specialisation and discrete division of work in a research process. It centres schema transforms and scripted crosswalks, and provides an alternative approach to existing data-centric methods.

Future areas for work include:

- Improving the script parser to better handle edge cases in source data terms (special, non-printing characters, as one example),
- Reviewing alternatives to Ray and Modin to improve speed and resilience in transforming extremely large data sources,
- Developing methods for integrating linked data into schemas so as to support, as example, externally-maintained ontologies for categorical terms,
- Improving the user-interface and contextual help for the web application,
- Offering a public repository of data sources and validated crosswalks.

## 6  ACKNOWLEDGEMENTS

whyqd received funding from the European Union's Horizon 2020 research and innovation programme under grant agreement No 101017536. Technical development support is from EOSC Future through the RDA Open Call mechanism, based on evaluations of external, independent experts.